\documentclass[3p]{elsarticle}

\usepackage[english]{babel}

\usepackage{graphicx}
\usepackage{dcolumn}
\usepackage{bm}

\usepackage{amsmath}
\usepackage{amssymb}

\usepackage{amsthm}
\newtheorem{lemma}{Lemma}
\newtheorem{definition}{Definition}

\usepackage{hyperref}

\begin{document}
\selectlanguage{english}

\title{On the reduced dynamics of \\a subset of interacting bosonic particles}

\author[fr,ino,inrim]{Manuel Gessner}
\ead{manuel.gessner@ino.it}
\author[fr,ox]{Andreas Buchleitner}%
\ead{a.buchleitner@physik.uni-freiburg.de}

\address[fr]{Physikalisches Institut, Albert-Ludwigs-Universit\"at Freiburg, Hermann-Herder-Stra\ss e 3, 79104 Freiburg, Germany}
\address[ino]{QSTAR, INO-CNR and LENS, Largo Enrico Fermi 2, I-50125 Firenze, Italy}
\address[inrim]{Istituto Nazionale di Ricerca Metrologica, Strada delle Cacce 91, I-10135 Torino, Italy}
\address[ox]{Keble College, University of Oxford, Oxford OX1 3PG, UK}

\date{\today}

\begin{abstract}
The quantum dynamics of a subset of interacting bosons in a subspace of fixed particle number is described in terms of symmetrized many-particle states. A suitable partial trace operation over the von Neumann equation of an $N$-particle system produces a hierarchical expansion for the subdynamics of $M\leq N$ particles. Truncating this hierarchy with a pure product state ansatz yields the general, nonlinear coherent mean-field equation of motion. In the special case of a contact interaction potential, this reproduces the Gross-Pitaevskii equation. To account for incoherent effects on top of the mean-field evolution, we discuss possible extensions towards a second-order perturbation theory that accounts for interaction-induced decoherence in form of a nonlinear Lindblad-type master equation.
\end{abstract}

\maketitle

\section{Introduction}
The effective quantum evolution of some pre-defined subsystem coupled to environmental degrees of freedom, realized by, e.g., a large number of particles, is often conveniently described by the theory of open quantum systems as formulated for clearly distinct system and environment \cite{Davies1976,Cohen-Tannoudji1992,AlickiLendi2007,BreuerPetruccione2006}. This is no longer true when the effective dynamics of a subset of identical particles is of interest. The natural decomposition into system and environment degrees of freedom which the standard open-system treatment relies upon is then unavailable and the \mbox{(anti-)}symmetrization postulate suggests that the effective dynamics of a subset of particles cannot be inferred without knowledge of the dynamics of the entire interacting system.

In stark contrast to the difficulty of theoretically predicting the subdynamics of identical particles stands the established and rather straight-forward experimental access to, e.g., single-particle observables of a many-body quantum system. In experiments with cold atoms, for instance, the dynamics of the average momentum distribution,
\begin{align}\label{eq.averagemom}
p_{\mathbf{k}}(t)=\mathrm{Tr}^{(N)}\{a^{\dagger}_{\mathbf{k}}a_{\mathbf{k}}\rho^{(N)}(t)\},
\end{align}
of an $N$-body system is easily measured \cite{Morsch,Meinert2014}. In~(\ref{eq.averagemom}), $a^{\dagger}_{\mathbf{k}}$ creates a particle with momentum $\mathbf{k}$, $\rho^{(N)}(t)$ denotes the quantum state of the full $N$-particle system at time $t$, and the trace $\mathrm{Tr}^{(N)}$ is performed over a complete basis of $N$-particle states. The probabilities $p_{\mathbf{k}}(t)$, in fact, define the diagonal elements of a single-particle density matrix, expressed in a basis of single-particle momentum eigenstates. Knowledge of the evolution of the single-particle density matrix would suffice to predict the expectation values of arbitrary single-particle operators, without necessarily knowing the evolution of the full $N$-particle density matrix. The same is true, e.g., for correlation functions of $M<N$ particles, given in terms of an $M$-particle density matrix. A description of the subdynamics of general $M$-particle density operators ($M\leq N$) in closed form therefore represents a fundamental problem and a long-standing challenge of mathematical and theoretical physics, with immediate relevance for the fields of atomic and solid-state physics, as well as quantum chemistry \cite{RevModPhys.34.694,Bogoliubov,Bogoliubov2,DavidsonBook,RDMBook,Roepke2013,PhysRevA.91.023412,Gertjerenken2015}.

Existing microscopic descriptions of the subdynamics of \textit{bosonic} particles are based on mean-field expansions, which are well-justified for modeling Bose-Einstein condensates \cite{PitS03}, i.e., when the  total many-body system is well described by a macroscopically occupied pure quantum state. If, additionally, the particles' interaction is limited to a contact potential, the dynamics of the single-particle state becomes nonlinear and is described by the Gross-Pitaevskii equation \cite{Gross1961,Pitaevskii1961}. While the Gross-Pitaevskii equation has been extremely successful in predicting the dynamics of ultracold quantum gases, it is unable to describe the effects of decoherence, due to the implicit assumption of a pure state for the subsystem at all times.

A more general picture is given by the Bogoliubov-Born-Green-Kirkwood-Young (BBGKY) hierarchy \cite{Yvon1935,Bogoliubov1946,Born1946,Kirkwood1946}, which describes the dynamics of the $M$-particle reduced density operator in the presence of arbitrary pairwise interactions in terms of the $(M+1)$-particle density operator \cite{Bogoliubov,Roepke2013}. For a closed description of the $M$-particle subdynamics, efficient and appropriate approximations for the truncation of the resulting hierarchical expansion must be found, and a general recipe to achieve this is not available. 

Presently there is no microscopic open-system theory for the incoherent subdynamics of identical particles. Interaction-induced decoherence of single-particle observables, such as the average momentum~(\ref{eq.averagemom}), however, was predicted by numerical studies \cite{BucK03,PonomarevDiss,Venzl2011} to manifest in decaying Bloch oscillations, and traced back to chaotic spectral statistics \cite{PhysRevE.68.056213}. This prediction was recently confirmed \cite{Meinert2014} in cold-atom experiments. 

In this article, we consider a closed system of a fixed number $N$ of bosonic particles, subject to pairwise interactions. We formally derive the equation of motion for the reduced density operator which predicts the quantum mechanical expectation values of any $M$-particle observable, with $M\leq N$. To do so we employ a description in terms of symmetrized states constructed on a subspace of fixed particle number. Our approach is based on extensions of concepts from open-system theory, such as the partial trace operation, to the considered scenario of indistinguishable particles.

We first show that by performing an appropriately defined partial trace operation on the von Neumann equation of the full many-body system, we recover a BBGKY-type hierarchy of equations of motion, which expresses the dynamics of the $M$-particle density operator as a function of the $(M+1)$-particle density matrix. This hierarchy can be truncated by a mean-field-type pure product state ansatz, where the contributions of higher particle numbers generate nonlinearities. The resulting equation of motion is shown to reproduce the Gross-Pitaevskii equation, in the case of a contact interaction potential.

In order to obtain an intrinsically incoherent equation of motion, we perform a second-order expansion of the full many-body von Neumann equation in terms of the particle interaction. The above partial trace then yields an equation of motion with a characteristic Lindblad-type structure, including dissipative terms which are derived microscopically. Again, the consequent hierarchical structure can be formally truncated with a pure product state ansatz, which leads to additional nonlinearities. In the limit of low densities, we discuss further approximations to obtain a dissipative mean-field equation, which extends the BBGKY-type hierarchy by additional, dissipative terms.

Nonlinear equations and hierarchical structures of dynamical equations are commonly encountered in many-body theory when dealing with bosonic particles in a mean-field approximation \cite{Bogoliubov,Bogoliubov2,Roepke2013,PitS03}. A description of decoherence, however, is usually only included as a result of an external bath which is clearly distinguishable from the system of interest \cite{Davies1976,Cohen-Tannoudji1992,AlickiLendi2007,BreuerPetruccione2006,DissBH}. Our approach outlines a path to obtaining a microscopic description of decoherence which arises due to the interaction of the particles among themselves, by combining a perturbative description of particle interactions with mean-field techniques.

\section{Mathematical framework}
\label{sec.generalformlism}
In this section we develop the mathematical framework of symmetrized quantum states of fixed particle numbers, which we will employ throughout this paper. Based on the bosonic creation and annihilation operators, associated with the modes $i$ and $j$,
\begin{align}\label{eq.ccrsingle}
[a_i,a^{\dagger}_j]=\delta_{ij},\qquad [a^{\dagger}_i,a^{\dagger}_j]=0,\qquad [a_i,a_j]=0,\qquad\forall i,j,
\end{align}
we construct a basis of \textit{symmetrized} $N$-particle states
\begin{align}\label{eq.creatvacsym}
|\varphi_{i_1}\dots\varphi_{i_N}\rangle=\frac{1}{\sqrt{N!}}a^{\dagger}_{i_1}\cdots a^{\dagger}_{i_N}|0\rangle,
\end{align}
where $|0\rangle$ denotes the vacuum state of no particles, and $\{|\varphi_i\rangle\}$ denotes the eigenbasis of an arbitrary single-particle observable. The completeness relation in the $N$-particle space then reads
\begin{align}\label{eq.completenessrelation}
\sum_{i_1\dots i_N}|\varphi_{i_1}\dots\varphi_{i_N}\rangle\langle\varphi_{i_1}\dots\varphi_{i_N}|=\mathbb{I}^{(N)},
\end{align}
where henceforth the superscript $(N)$ denotes the $N$-particle space. The sums in the above expression extend over the entire set of basis states, and in the case of a continuous basis, they are replaced by an integral. Using the completeness of the symmetrized states, we can express arbitrary $N$-particle operators as
\begin{align}\label{eq.npartoperator}
A^{(N)}=\frac{1}{N!}\sum_{\substack{i_1\dots i_N\\j_1 \dots j_N}}A^{(N)}_{i_1 \dots i_N;j_1 \dots j_N}a^{\dagger}_{i_1}\dots a^{\dagger}_{i_N}a_{j_1}\dots a_{j_N},
\end{align}
where
\begin{align}\label{eq.nparticleoperator}
A^{(N)}_{i_1\dots i_N;j_1\dots j_N}&=\langle\varphi_{i_1}\dots\varphi_{i_N}|A^{(N)}|\varphi_{j_1}\dots\varphi_{j_N}\rangle.
\end{align}
Notice that these symmetric constructions are valid only for bosonic systems and for fermions sign changes would occur due to the reordering of fields. In writing Eq.~(\ref{eq.npartoperator}), we have omitted the vacuum projector $|0\rangle\langle 0|$ (which origins in Eq.~(\ref{eq.creatvacsym})), to allow for the determination of the expectation value of $A^{(N)}$ in a space of more than $N$ particles. This will become relevant later for the definition of reduced density operators.

The action of creation and annihilation operators $a^{\dagger}_k$ and $a_k$ on the symmetrized $N$-particle states is given by
\begin{align}\label{eq.creasym}
a^{\dagger}_k|\varphi_{i_1}\dots\varphi_{i_N}\rangle=\sqrt{N+1}|\varphi_k\varphi_{i_1}\dots\varphi_{i_N}\rangle,
\end{align}
and
\begin{align}\label{eq.annisym}
a_k|\varphi_{i_1}\dots\varphi_{i_N}\rangle=\frac{1}{\sqrt{N}}\sum_{j=1}^N\delta_{ki_j}|\varphi_{i_1}\dots\varphi_{i_{j-1}}\varphi_{i_{j+1}}\dots\varphi_{i_N}\rangle.
\end{align}
Finally, the scalar product of two symmetrized states follows as
\begin{align}
\langle\varphi_{i_1}\dots\varphi_{i_N}|\varphi_{j_1}\dots\varphi_{j_N}\rangle=\frac{1}{N!}\sum_{\sigma\in\mathcal{S}_N}\delta_{i_{1}j_{\sigma(1)}}\cdots\delta_{i_{N}j_{\sigma(N)}},
\end{align}
where $\sigma$ denotes a permutation taken from the symmetric group $\mathcal{S}_N$ of $N$ particles. The orthonormality condition can be expressed as
\begin{align}\label{eq.spsym}
\langle\varphi_{i_1}\dots\varphi_{i_N}|\varphi_{j_1}\dots\varphi_{j_N}\rangle=\delta_{i_1\dots i_N;j_1\dots j_N},
\end{align}
with the following convenient
\begin{definition}[Permutation-invariant Kronecker delta]\label{def.def1}
\begin{align}\label{eq.perminvdelta}
\delta_{i_1\dots i_N;j_1\dots j_N}:=\frac{1}{N!}\sum_{\sigma\in\mathcal{S}_N}\prod_{k=1}^N\delta_{i_kj_{\sigma(k)}}.
\end{align}
\end{definition}
Henceforth the semicolon separates two independent sets of indices. 

Obviously, $\delta_{i_1\dots i_N;j_1\dots j_N}=\delta_{j_1\dots j_N;i_1\dots i_N}$. If one of the sets $\{i_1\dots i_N\}$ and $\{j_1\dots j_N\}$ is invariant under all permutations, the permutation-invariant Kronecker delta reduces to the simple product
\begin{align}
\delta_{i_1\dots i_N;j_1\dots j_N}=\prod_{k=1}^N\delta_{i_kj_{k}}.
\end{align}
In the following we will frequently encounter expressions whose indices are symmetric under permutations, due to bosonic quantum statistics. In this context, the permutation-invariant Kronecker delta is particularly practical. To see this, let $f(i_1\dots i_N)$ be a function which is invariant under permutations within the set of indices $i_1\dots i_N$,
\begin{align}\label{eq.perminvariant}
f(i_1\dots i_N)=f(i_{\sigma(1)}\dots i_{\sigma(N)}),
\end{align}
for all $\sigma\in\mathcal{S}_N$. Then the following holds
\begin{align}\label{eq.krondeltuse}
\sum_{i_1\dots i_N}f(i_1\dots i_N)\delta_{i_1\dots i_N;j_1\dots j_N}=\frac{1}{N!}\sum_{\sigma\in\mathcal{S}_N}f(j_{\sigma(1)}\dots j_{\sigma(N)})=f(j_1\dots j_N).
\end{align}
Note that the permutation-invariant Kronecker delta itself fulfills property~(\ref{eq.perminvariant}).

We conclude this section by generalizing Eqs.~(\ref{eq.creasym}) and~(\ref{eq.annisym}) to arbitrary numbers $M$ of creation and annihilation operators. From the obtained expressions we infer further rules for the convenient treatment of symmetrized states, which will become useful in later sections. First, we generalize Eq.~(\ref{eq.annisym}) to
\begin{lemma}\label{lm.lemma3a}
\begin{align}\label{eq.lemma1}
&a_{i_1}\dots a_{i_M}|\varphi_{j_1}\dots\varphi_{j_N}\rangle\notag\\
&=\sqrt{\frac{(N-M)!}{N!}}M!\sum_{1\leq\alpha_1<\dots<\alpha_M\leq N}\delta_{i_1\dots i_M;j_{\alpha_1}\dots j_{\alpha_M}}|\varphi_{\{j_1\dots j_N\}\backslash\{j_{\alpha_1}\dots j_{\alpha_M}\}}\rangle,
\end{align}
\textnormal{where we introduced the shorthand notation $|\varphi_{\{j_1\dots j_N\}}\rangle=|\varphi_{j_1}\dots\varphi_{j_N}\rangle$. \\The set $\{j_1\dots j_N\}\backslash\{j_{\alpha_1}\dots j_{\alpha_M}\}$ is obtained by removing the elements $\{j_{\alpha_1}\dots j_{\alpha_M}\}$ from the set $\{j_1\dots j_N\}$. The vacuum state is included as $|\varphi_{\{\}}\rangle=|0\rangle$.}
\begin{proof} We employ the canonical commutation relations~(\ref{eq.ccrsingle}), and use the permutation-invariant Kronecker delta~(\ref{eq.perminvdelta}), to obtain:
\begin{align}
&a_{i_1}\dots a_{i_M}|\varphi_{j_1}\dots\varphi_{j_N}\rangle\notag\\
&=\frac{1}{\sqrt{N!}}a_{i_M}\dots a_{i_1}a^{\dagger}_{j_1}\dots a^{\dagger}_{j_N}|0\rangle\notag\\
&=\frac{1}{\sqrt{N!}}\sum_{\alpha_1\in I}\dots\sum_{\alpha_M\in I\backslash\{\alpha_1\dots \alpha_{M-1}\}}\delta_{i_1j_{\alpha_1}}\dots\delta_{i_Mj_{\alpha_M}}\prod_{l\in I\backslash\{\alpha_1\dots \alpha_M\}}a^{\dagger}_{j_l}|0\rangle\notag\\
&=\frac{M!}{\sqrt{N!}}\sum_{1\leq \alpha_1<\dots<\alpha_M\leq N}\delta_{i_1\dots i_M;j_{\alpha_1}\dots j_{\alpha_M}}\prod_{l\in I\backslash\{\alpha_1\dots \alpha_M\}}a^{\dagger}_{j_l}|0\rangle\notag\\
&=\sqrt{\frac{(N-M)!}{N!}}M!\sum_{1\leq\alpha_1<\dots<\alpha_M\leq N}\delta_{i_1\dots i_M;j_{\alpha_1}\dots j_{\alpha_M}}|\varphi_{\{j_1\dots j_N\}\backslash\{j_{\alpha_1}\dots j_{\alpha_M}\}}\rangle,
\end{align}
with $I=\{1,\dots, N\}$.
\end{proof}
\end{lemma}
For $M=1$ Eq.~(\ref{eq.lemma1}) from \textit{Lemma \ref{lm.lemma3a}} indeed reduces to Eq.~(\ref{eq.annisym}). Furthermore, when $M=N$, we obtain
\begin{align}\label{eq.lemma3}
a_{i_1}\dots a_{i_N}|\varphi_{j_1}\dots\varphi_{j_N}\rangle=\delta_{i_1\dots i_N;j_1\dots j_N}\sqrt{N!}|0\rangle,
\end{align}
which is compatible with Eq.~(\ref{eq.spsym}):
\begin{align}\label{eq.normalization}
\langle\varphi_{i_1}\dots\varphi_{i_N}|\varphi_{j_1}\dots\varphi_{j_N}\rangle&=\frac{1}{\sqrt{N!}}\langle0|a_{i_1}\dots a_{i_N}|\varphi_{j_1}\dots\varphi_{j_N}\rangle=\delta_{i_1\dots i_N;j_1\dots j_N}.
\end{align}
We further generalize Eq.~(\ref{eq.creasym}) to
\begin{align}\label{eq.23}
a^{\dagger}_{i_1}\dots a^{\dagger}_{i_M}|\varphi_{j_1}\dots\varphi_{j_N}\rangle=\sqrt{\frac{(N+M)!}{N!}}|\varphi_{i_1}\dots\varphi_{i_M}\varphi_{j_1}\dots\varphi_{j_N}\rangle,
\end{align}
which leads us to
\begin{align}\label{eq.sumexpression}
\sum_{i_1\dots i_M}a^{\dagger}_{i_1}\dots a^{\dagger}_{i_M}a_{i_1}\dots a_{i_M}|\varphi_{j_1}\dots\varphi_{j_N}\rangle=\frac{N!}{(N-M)!}|\varphi_{j_1}\dots\varphi_{j_N}\rangle.
\end{align}
This follows immediately from \textit{Lemma \ref{lm.lemma3a}} and Eq.~(\ref{eq.23}). For $M=1$, expression~(\ref{eq.sumexpression}) yields the number operator with eigenvalue $N$.

Using the cyclicity of the trace together with Eqs.~(\ref{eq.lemma3}) and (\ref{eq.krondeltuse}), we obtain, for an arbitrary operator $X$:
\begin{align}\label{eq.lemma6}
\mathrm{Tr}^{(N)}\{a_{i_1}\dots a_{i_N} a^{\dagger}_{j_1}\dots a^{\dagger}_{j_N}X\}&=\sum_{\alpha_1\dots \alpha_N}\langle\varphi_{\alpha_1}\dots\varphi_{\alpha_N}|a^{\dagger}_{j_1}\dots a^{\dagger}_{j_N}Xa_{i_1}\dots a_{i_N}|\varphi_{\alpha_1}\dots\varphi_{\alpha_N}\rangle\notag\\
&=N!\sum_{\alpha_1\dots \alpha_N}\delta_{\alpha_1\dots \alpha_N;j_1\dots j_N}\delta_{i_1\dots i_N;\alpha_1\dots \alpha_N}\langle 0|X|0\rangle\notag\\
&=N!\delta_{i_1\dots i_N;j_1\dots j_N}\langle 0|X|0\rangle.
\end{align}
Similarly,
\begin{align}\label{eq.lemma7}
\mathrm{Tr}^{(N)}\{a^{\dagger}_{j_1}\dots a^{\dagger}_{j_N}a_{i_1}\dots a_{i_N}X\}
&=\sum_{\alpha_1\dots \alpha_N}\langle\varphi_{\alpha_1}\dots\varphi_{\alpha_N}|Xa^{\dagger}_{j_1}\dots a^{\dagger}_{j_N}a_{i_1}\dots a_{i_N}|\varphi_{\alpha_1}\dots\varphi_{\alpha_N}\rangle\notag\\
&=N!\sum_{\alpha_1\dots \alpha_N}\langle\varphi_{\alpha_1}\dots\varphi_{\alpha_N}|X|\varphi_{j_1}\dots\varphi_{j_N}\rangle\delta_{i_1\dots i_N;\alpha_1\dots \alpha_N}\notag\\
&=N!\langle\varphi_{i_1}\dots\varphi_{i_N}|X|\varphi_{j_1}\dots\varphi_{j_N}\rangle.
\end{align}

Next, we apply these rules to determine matrix elements and traces of arbitrary $M$-particle operators $A^{(M)}$ in an $N$-particle space, which will appear frequently in our subsequent description of bosonic subsystems. An important tool in this context will be the following \textit{Lemma}, which we obtain with the help of \textit{Lemma \ref{lm.lemma3a}}:
\begin{lemma}\label{lm.lemma3c}
\begin{align}
&\langle\varphi_{i_1}\dots\varphi_{i_N}|a^{\dagger}_{k_1}\dots a^{\dagger}_{k_M}a_{l_1}\dots a_{l_M}|\varphi_{j_1}\dots\varphi_{j_N}\rangle\notag\\
&=\frac{(N-M)!}{N!}M!^2\sum_{\substack{1\leq\alpha_1<\dots<\alpha_M\leq N\\1\leq\beta_1<\dots<\beta_M\leq N}}\delta_{k_1\dots k_M;i_{\alpha_1}\dots i_{\alpha_M}}\delta_{l_1\dots l_M;j_{\beta_1}\dots j_{\beta_M}}\notag\\&\hspace{5cm}\times
\delta_{\{i_1\dots i_N\}\backslash\{i_{\alpha_1}\dots i_{\alpha_M}\};\{j_1\dots j_N\}\backslash\{j_{\beta_1}\dots j_{\beta_M}\}}.
\end{align}
\end{lemma}
This yields
\begin{align}\label{eq.npartm}
&\quad\langle\varphi_{i_1}\dots\varphi_{i_N}|A^{(M)}|\varphi_{j_1}\dots\varphi_{j_N}\rangle\\
&=\frac{1}{M!}\sum_{\substack{k_1\dots k_M\\ l_1\dots l_M}}A^{(M)}_{k_1\dots k_M;l_1\dots l_M}\langle\varphi_{i_1}\dots\varphi_{i_N}|a^{\dagger}_{k_1}\dots a^{\dagger}_{k_M}a_{l_1}\dots a_{l_M}|\varphi_{j_1}\dots\varphi_{j_N}\rangle\notag\\
&=\binom{N}{M}^{-1}\sum_{\substack{1\leq\alpha_1<\dots<\alpha_M\leq N\\1\leq\beta_1<\dots<\beta_M\leq N}}A^{(M)}_{i_{\alpha_1}\dots i_{\alpha_M};j_{\beta_1}\dots j_{\beta_M}}\delta_{\{i_1\dots i_N\}\backslash\{i_{\alpha_1}\dots  i_{\alpha_M}\};\{j_1\dots j_N\}\backslash\{j_{\beta_1}\dots  j_{\beta_M}\}}.\notag
\end{align}

For the $N$-particle trace of $M$-particle operators we will encounter expressions of the following type:
\begin{align}\label{eq.nparttracewithm}
&\qquad\mathrm{Tr}^{(N)}\{a^{\dagger}_{n_1}\dots a^{\dagger}_{n_M}a_{m_1}\dots a_{m_M}X\}\notag\\
&=\sum_{i_1\dots i_N}\langle\varphi_{i_1}\dots\varphi_{i_N}|a^{\dagger}_{n_1}\dots a^{\dagger}_{n_M}a_{m_1}\dots a_{m_M}X|\varphi_{i_1}\dots\varphi_{i_N}\rangle\notag\\
&=M!\sum_{i_1\dots i_N}\sum_{1\leq\alpha_1<\dots<\alpha_M\leq N}\delta_{i_{\alpha_1}\dots  i_{\alpha_M};n_1\dots n_M}\langle\varphi_{\{i_1\dots  i_N\}\backslash \{i_{\alpha_1}\dots  i_{\alpha_M}\}\cup\{{m_1}\dots {m_M}\}}|X|\varphi_{i_1}\dots\varphi_{i_N}\rangle\notag\\
&=\frac{N!}{(N-M)!}\sum_{i_{M+1}\dots i_N}\langle\varphi_{m_1}\dots\varphi_{m_M}\varphi_{i_{M+1}}\dots\varphi_{i_N}|X|\varphi_{n_1}\dots\varphi_{n_M}\varphi_{i_{M+1}}\dots\varphi_{i_N}\rangle.
\end{align}
After the second equality above, the set $\{i_1\dots  i_N\}\backslash \{i_{\alpha_1}\dots  i_{\alpha_M}\}\cup\{{m_1}\dots {m_M}\}$ appears, which is obtained from the set $\{i_1\dots  i_N\}$ by replacing the elements $\{i_{\alpha_1}\dots  i_{\alpha_M}\}$ with $\{{m_1}\dots {m_M}\}$. The order of the indices $i_1\dots i_N$ can be changed by relabeling, which yields a sum over $\binom{N}{M}$ identical terms in the third step.

\section{Density operators and expectation values}
A special case of an $N$-particle operator is the density operator
\begin{align}
\rho^{(N)}=\frac{1}{N!}\sum_{\substack{i_1\dots i_N\\j_1\dots j_N}}\rho^{(N)}_{i_1 \dots i_N;j_1 \dots j_N}a^{\dagger}_{i_1}\dots a^{\dagger}_{i_N}a_{j_1}\dots a_{j_N},
\end{align}
which describes the quantum state of an $N$-particle system, and is subject to the normalization condition $\mathrm{Tr}^{(N)}\{\rho^{(N)}\}=\sum_{i_1\dots i_N}\rho^{(N)}_{i_1 \dots i_N;i_1\dots i_N}=1$. The quantum mechanical expectation value for measurements of the observable $A^{(N)}$ is then obtained, with Eq.~(\ref{eq.nparticleoperator}), as
\begin{align}\label{eq.nparticleexpectationvalue}
\mathrm{Tr}^{(N)}\{\rho^{(N)}A^{(N)}\}&=\frac{1}{N!^2}\sum_{\substack{i_1\dots i_N\\j_1\dots j_N}}\sum_{\substack{n_1\dots n_N\\m_1\dots m_N}}\rho^{(N)}_{i_1 \dots i_N;j_1 \dots j_N}A^{(N)}_{n_1 \dots n_N;m_1 \dots m_N}\notag\\&\hspace{1cm}\times\mathrm{Tr}^{(N)}\{a^{\dagger}_{i_1}\dots a^{\dagger}_{i_N}a_{j_1}\dots a_{j_N}a^{\dagger}_{n_1}\dots a^{\dagger}_{n_N}a_{m_1}\dots a_{m_N}\}\notag\\
&=\sum_{\substack{i_1\dots i_N\\j_1\dots j_N}}\sum_{\substack{n_1\dots n_N\\m_1\dots m_N}}\rho^{(N)}_{i_1 \dots i_N;j_1 \dots j_N}A^{(N)}_{n_1 \dots n_N;m_1 \dots m_N}
\delta_{m_1\dots m_N;i_1\dots i_N}\delta_{j_1\dots j_N;n_1\dots n_N}\notag\\
&=\sum_{i_1\dots i_N}\sum_{n_1\dots n_N}\rho^{(N)}_{i_1\dots i_N;n_1 \dots n_N}A^{(N)}_{n_1 \dots n_N;i_1\dots i_N},
\end{align}
where we used Eqs.~(\ref{eq.normalization}) and (\ref{eq.lemma6}).

To introduce the reduced density operator, we consider the expectation value of an $M$-particle operator in an $N$-particle system ($M\leq N$):
\begin{align}\label{eq.mreduceddensopexp}
\mathrm{Tr}^{(N)}\{\rho^{(N)}A^{(M)}\}&=\frac{1}{M!}\sum_{\substack{n_1\dots n_M\\m_1\dots m_M}}\mathrm{Tr}^{(N)}\{a_{n_1}^{\dagger}\dots a_{n_M}^{\dagger}a_{m_1}\dots a_{m_M}\rho^{(N)}\}A^{(M)}_{n_1 \dots n_M;m_1 \dots m_M}\notag\\
&=\binom{N}{M}\sum_{\substack{n_1\dots n_M\\m_1\dots m_M}}\rho^{(M)}_{m_1 \dots m_M;n_1 \dots n_M}A^{(M)}_{n_1 \dots n_M;m_1 \dots m_M}\notag\\
&=\binom{N}{M}\mathrm{Tr}^{(M)}\{\rho^{(M)}A^{(M)}\},
\end{align}
where we used Eq.~(\ref{eq.nparticleexpectationvalue}) in the last step. The expectation value is therefore fully determined by the \textit{$M$-particle reduced density operator of an $N$-particle system}\cite{Husimi,PhysRev.97.1474,PhysRev.100.1579,terHaar1961,RevModPhys.34.694,Bogoliubov,DavidsonBook,RDMBook}, with matrix elements, which, by comparison of the first and second line in Eq.~(\ref{eq.mreduceddensopexp}), are given by
\begin{align}\label{eq.reducedelements}
\rho^{(M)}_{i_1\dots i_M;j_1\dots j_M}&=\frac{(N-M)!}{N!}\mathrm{Tr}^{(N)}\{a^{\dagger}_{j_1}\dots a^{\dagger}_{j_M}a_{i_1}\dots a_{i_M}\rho^{(N)}\}\\
&=\sum_{n_{M+1}\dots n_N}\langle \varphi_{i_1}\dots\varphi_{i_M}\varphi_{n_{M+1}}\dots\varphi_{n_N}|\rho^{(N)}|\varphi_{j_1}\dots\varphi_{j_M}\varphi_{n_{M+1}}\dots\varphi_{n_M}\rangle\notag,
\end{align}
where we used Eq.~(\ref{eq.nparttracewithm}) in the last step. The matrix elements~(\ref{eq.reducedelements}) define a reduced density operator and contain all the necessary information for the description of a subset of $M$ bosonic particles---such as, e.g., the average momentum, Eq.~(\ref{eq.averagemom}). This can be interpreted as the direct analog of the reduced density matrices known from the conventional treatment of composite quantum systems, i.e., systems of various distinguishable degrees of freedom. These reduced density matrices completely determine all measurement results of operators that restrict on a subset of these degrees of freedom and, thus, are central to the theory of open quantum systems \cite{Davies1976,AlickiLendi2007,BreuerPetruccione2006}.

The reduced density operator for $M$ indistinguishable bosons can be obtained from the full density operator $\rho^{(N)}$ via a suitable partial trace operation as $\rho^{(M)}=\mathrm{Tr}^{(N)}_{N-M}\{\rho^{(N)}\}$. The corresponding partial trace operation over $N-M$ out of $N$ identical bosonic particles is defined as
\begin{align}\label{eq.reduced}
\mathrm{Tr}^{(N)}_{N-M}\{X^{(N)}\}&=\frac{(N-M)!}{N!M!}\sum_{\substack{n_1\dots n_M\\m_1\dots m_M}}\mathrm{Tr}^{(N)}\{a_{n_1}^{\dagger}\dots a_{n_M}^{\dagger}a_{m_1}\dots a_{m_M}X^{(N)}\}\notag\\&\hspace{3.5cm}\times a_{m_1}^{\dagger}\dots a_{m_M}^{\dagger}a_{n_1}\dots a_{n_M},
\end{align}
where $X^{(N)}$ is an arbitrary $N$-particle operator. The above definition maps an $N$-particle operator onto an $M$-particle operator. Proper normalization \cite{TerHaar1960} of $\rho^{(M)}$ is provided by means of Eqs.~(\ref{eq.sumexpression}) and (\ref{eq.normalization}):
\begin{align}
\mathrm{Tr}^{(M)}\{\rho^{(M)}\}&=\frac{(N-M)!}{N!}\sum_{\substack{i_1\dots i_N\\j_1\dots j_N\\n_1\dots n_M}}\rho^{(N)}_{i_1 \dots i_N;j_1 \dots j_N}\langle\varphi_{j_1}\dots \varphi_{j_N}|a_{n_1}^{\dagger}\dots a_{n_M}^{\dagger}a_{n_1}\dots a_{n_M}|\varphi_{i_1}\dots\varphi_{i_N}\rangle\notag\\
&=\frac{(N-M)!}{N!}\frac{N!}{(N-M)!}\underbrace{\sum_{i_1\dots i_N}\rho^{(N)}_{i_1 \dots i_N;i_1 \dots i_N}}_{1}=1.
\end{align}

The above definition~(\ref{eq.reduced}) of the partial trace can be interpreted as a projection of $N$-particle operators onto the subspace spanned by $M$-particle states. This can be realized using the completeness~(\ref{eq.completenessrelation}) of the symmetrized states
\begin{align}\label{eq.hspacetruncation}
\mathbb{I}^{(M)}X^{(N)}\mathbb{I}^{(M)}&=\sum_{\substack{n_1\dots n_M\\m_1\dots m_M}}|\varphi_{n_1}\dots \varphi_{n_M}\rangle\langle \varphi_{n_1}\dots \varphi_{n_M}|X^{(N)}|\varphi_{m_1}\dots \varphi_{m_M}\rangle\langle \varphi_{m_1}\dots \varphi_{m_M}|\notag\\
&=\frac{1}{M!^2}\sum_{\substack{n_1\dots n_M\\m_1\dots m_M}}\mathrm{Tr}^{(N)}\left\{a^{\dagger}_{m_1}\dots a^{\dagger}_{m_M}a_{n_1}\dots a_{n_M}X^{(N)}\right\} a^{\dagger}_{n_1}\dots a^{\dagger}_{n_M}a_{m_1}\dots a_{m_M}\notag\\
&=\binom{N}{M}\mathrm{Tr}^{(N)}_{N-M}\{X^{(N)}\},
\end{align}
where, as in Eq.~(\ref{eq.npartoperator}), we omitted the projector on the vacuum state in the second step.

\section{Hierarchical expansion of the reduced bosonic dynamics}\label{sec.bbgky}
The many-body dynamics of an $N$-particle system is governed by the von Neumann equation
\begin{align}\label{eq.vnspte}
\frac{\partial}{\partial t}\rho^{(N)}(t)=-\frac{i}{\hbar}\left[H,\rho^{(N)}(t)\right].
\end{align}
Here, we consider second-quantized Hamiltonians $H$ which decompose into a single-particle term $H^{(1)}$, consisting of kinetic terms and an external potential, and a two-particle interaction term $H^{(2)}$:
\begin{align}\label{eq.generalmanybodyH}
H=H^{(1)}+H^{(2)}=\sum_{ij}H^{(1)}_{ij}a^{\dagger}_ia_j+\frac{1}{2}\sum_{ijkl}H^{(2)}_{ij;kl}a^{\dagger}_ia^{\dagger}_ja_ka_l.
\end{align}
The single-particle term $H^{(1)}$ can be eliminated by transforming to the interaction picture with $U_1(t)=\exp(-iH^{(1)}t/\hbar)$. The dynamics is then described equivalently by
\begin{align}\label{eq.vNinter}
\frac{\partial}{\partial t}\rho^{(N)}_I(t)=-\frac{i}{\hbar}[H^{(2)}_I(t),\rho_I^{(N)}(t)],
\end{align}
with $\rho^{(N)}_I(t)=U^{\dagger}_1(t)\rho^{(N)}(t)U_1(t)$ and $H^{(2)}_I(t)=U^{\dagger}_1(t)H^{(2)}U_1(t)$. We will make use of the interaction picture later in this manuscript.

Notice that on the left-hand side of the von Neumann equation~(\ref{eq.vnspte}), the time derivative of an $N$-particle operator appears. This derivative can, in principle, contain contributions from the many-particle subspace with $N+1$ particles, generated by the commutator on the right-hand side through terms, such as $H^{(1)}\rho^{(N)}$ and $H^{(2)}\rho^{(N)}$.\footnote{In fact, also operators on $N+2$ particles emerge, but eventually disappear due to the commutator.} These are, however, only relevant for the infinitesimal propagation described by the differential equation~(\ref{eq.vnspte}), since Hamiltonians of the form~(\ref{eq.generalmanybodyH}) always commute with the particle number operator $N^{(1)}=\sum_ia^{\dagger}_ia_i$. Hence, the state at the outcome of the time evolution, i.e., the solution to Eq.~(\ref{eq.vnspte}), will always be an $N$-particle operator.

\subsection{Reduced dynamics in the $M$-particle subspace}
To describe the dynamics of the reduced $M$-particle density operator, we perform the partial trace operation, as defined in Eq.~(\ref{eq.reduced}), on both sides of the von Neumann equation~(\ref{eq.vnspte}), leading to
\begin{align}\label{eq.startingequationbbgky}
i\hbar\frac{\partial}{\partial t}\rho^{(M)}(t)=\mathrm{Tr}^{(N)}_{N-M}\left\{\left[H^{(1)}+H^{(2)},\rho^{(N)}(t)\right]\right\}.
\end{align}
By construction, the partial trace operation $\mathrm{Tr}^{(N)}_{N-M}$ treats its argument as an $N$-particle operator and projects it onto the $M$-particle subspace. Thus, in Eq.~(\ref{eq.startingequationbbgky})
\begin{itemize}
\item[(i)] only $N$-particle contributions to the right-hand side of Eq.~(\ref{eq.vnspte}) are considered, and 
\item[(ii)] only $M$-particle contributions to $\partial_t\rho^{(M)}(t)$ are generated.
\end{itemize}
As was discussed before, such restrictions do not apply to the full von Neumann equation~(\ref{eq.vnspte}). In this paper, our aim is to find an effective description for the dynamics of $\rho^{(M)}$, based on the description within the $M$-particle subspace provided by Eq.~(\ref{eq.startingequationbbgky}).

To evaluate the right-hand side of Eq.~(\ref{eq.startingequationbbgky}), we treat the single-particle and two-particle contributions to $H$ separately. With Eq.~(\ref{eq.reduced}) and \textit{Lemma~\ref{lm.lemma3a}}, we obtain for the single-particle part
\begin{align}
&\quad\mathrm{Tr}^{(N)}_{N-M}\left\{\left[H^{(1)},\rho^{(N)}(t)\right]\right\}\notag\\
&=\frac{(N-M)!}{N!M!}\notag\\&\quad\times\sum_{\substack{n_1 \dots n_M\\m_1 \dots m_M\\k_1 \dots k_N\\l_1\dots l_N}}
\left[\langle \varphi_{l_1}\dots \varphi_{l_N}|a^{\dagger}_{n_1}\dots a^{\dagger}_{n_M}a_{m_1}\dots a_{m_M}H^{(1)}|\varphi_{k_1}\dots\varphi_{k_N}\rangle\rho^{(N)}_{k_1\dots k_N;l_1\dots l_N}(t)\right.\notag\\
&\left.\hspace{2cm}-\langle \varphi_{l_1}\dots \varphi_{l_N}|H^{(1)}a^{\dagger}_{n_1}\dots a^{\dagger}_{n_M}a_{m_1}\dots a_{m_M}|\varphi_{k_1}\dots\varphi_{k_N}\rangle\rho^{(N)}_{k_1\dots k_N;l_1\dots l_N}(t)\right]\notag\\
&\hspace{2cm} \times a^{\dagger}_{m_1}\dots a^{\dagger}_{m_M}a_{n_1}\dots a_{n_M}.\notag\\
&=\frac{(N-M)!}{N!M!}\binom{N}{M}M!\sum_{\substack{n_1 \dots n_M\\m_1 \dots m_M}}\notag\\
&\quad\times\left[\sum_{\substack{k_1\dots k_N\\l_{M+1}\dots l_N}}\langle \varphi_{m_1}\dots\varphi_{m_M}\varphi_{l_{M+1}}\dots \varphi_{l_N}|H^{(1)}|\varphi_{k_1}\dots \varphi_{k_N}\rangle\rho^{(N)}_{k_1\dots k_N;n_1\dots n_Ml_{M+1}\dots l_N}(t)\right.\notag\\
&\left.\qquad-\sum_{\substack{k_1 \dots k_{M+1}\\l_{1}\dots l_N}}\langle \varphi_{l_1}\dots \varphi_{l_N}|H^{(1)}|\varphi_{n_1}\dots \varphi_{n_M} \varphi_{k_{M+1}}\dots \varphi_{k_N}\rangle\rho^{(N)}_{m_1\dots m_Mk_{M+1}\dots k_N;l_1\dots l_N}(t)\right]\notag\\
&\quad \times a^{\dagger}_{m_1}\dots a^{\dagger}_{m_M}a_{n_1}\dots a_{n_M}.
\end{align}
The required $N$-particle matrix elements of a single-particle operator are obtained from Eq.~(\ref{eq.npartm}) with $M=1$, where the label $1\leq\alpha\leq N$ appears. There are $M$ possibilities for the label $\alpha$ to be found within the subset $1,\dots,M$. In all these cases one index of $H^{(1)}$ is directly contracted with either the creation or the annihilation operators $a^{\dagger}_{m_1},\dots,a^{\dagger}_{m_M}$ and $a_{n_1},\dots, a_{n_M}$. In the remaining $N-M$ cases, both indices of $H^{(1)}$ are contracted with $\rho^{(N)}$. Explicit discrimination of these two cases leads to
\begin{align}
&\quad\mathrm{Tr}^{(N)}_{N-M}\left\{\left[H^{(1)},\rho^{(N)}(t)\right]\right\}\notag\\
&=\frac{(N-M)!}{N!M!}\binom{N}{M}M!\sum_{\substack{n_1 \dots n_M\\m_1 \dots m_M}}\notag\\
&\quad\times\left[M\left(\sum_{k_1}H^{(1)}_{m_1k_1}\rho^{(M)}_{k_1m_2\dots m_M;n_1\dots n_M}(t)-\sum_{l_1}H^{(1)}_{l_1n_1}\rho^{(M)}_{m_1\dots m_M;n_2\dots n_Ml_1}(t)\right)\right.\notag\\
&\qquad+(N-M)\left(\sum_{l_{M+1}}\sum_{k_1}H^{(1)}_{l_{M+1}k_1}\rho^{(M+1)}_{k_1m_1\dots m_M;n_1\dots n_Ml_{M+1}}(t)\right.\notag\\
&\hspace{3.2cm}-\left.\left.\sum_{l_1}\sum_{k_{M+1}}H^{(1)}_{l_1k_{M+1}}\rho^{(M+1)}_{m_1\dots m_Mk_{M+1};l_1n_1\dots n_M}(t)\right)\right] a^{\dagger}_{m_1}\dots a^{\dagger}_{m_M}a_{n_1}\dots a_{n_M}.
\end{align}
Since the matrix elements of $\rho^{(M+1)}$ do not depend on a specific order of the indices, we find by relabeling of the indices that the terms containing $\rho^{(M+1)}$ cancel. This finally yields
\begin{align}\label{eq.singleparticlebbgky}
&\quad\mathrm{Tr}^{(N)}_{N-M}\left\{\left[H^{(1)},\rho^{(N)}(t)\right]\right\}\notag\\
&=\frac{M}{M!}\sum_{\substack{n_1 \dots n_M\\m_1 \dots m_M\\k_1}}\left[H^{(1)}_{m_1k_1}\rho^{(M)}_{k_1m_2\dots m_M;n_1\dots n_M}(t)-H^{(1)}_{k_1n_1}\rho^{(M)}_{m_1\dots m_M;n_2\dots n_Mk_1}(t)\right] a^{\dagger}_{m_1}\dots a^{\dagger}_{m_M}a_{n_1}\dots a_{n_M}\notag\\
&=[H^{(1)},\rho^{(M)}(t)].
\end{align}
The factor $M$ is due to the representation of the single-particle operator $H^{(1)}$ in a basis of $M$-particle states. In general, we find for an arbitrary $S$-particle operators $J^{(S)}$,
\begin{align}\label{eq.ms1}
\langle \varphi_{n_1}\dots\varphi_{n_M}|J^{(S)}\rho^{(M)}|\varphi_{m_1}\dots\varphi_{m_M}\rangle=\binom{M}{S}\sum_{i_1\dots i_S}J^{(S)}_{n_1\dots n_S;i_1\dots i_S}\rho^{(M)}_{i_1\dots i_Sn_{S+1}\dots n_M;m_1\dots m_M},
\end{align}
and
\begin{align}\label{eq.ms2}
\langle \varphi_{n_1}\dots\varphi_{n_M}|\rho^{(M)}J^{(S)}|\varphi_{m_1}\dots\varphi_{m_M}\rangle=\binom{M}{S}\sum_{i_1\dots i_S}\rho^{(M)}_{n_1\dots n_M;i_1\dots i_Sm_{S+1}\dots m_M}J^{(S)}_{i_1\dots i_S;m_1\dots m_S}.
\end{align}
For $S=1$ we indeed obtain the prefactor $M$ which appeared in Eq.~(\ref{eq.singleparticlebbgky}). 

As expected, in the absence of interactions $H^{(2)}$, Eqs.~(\ref{eq.startingequationbbgky}) and~(\ref{eq.singleparticlebbgky}) describe the fully coherent subdynamics of $M\leq N$ particles. This observation is independent of the initial state $\rho^{(N)}$ and the correlations it may contain.

Next, we study the contribution of the two-particle interaction term $H^{(2)}$ to Eq.~(\ref{eq.startingequationbbgky}) for $M\geq2$. We obtain, again using Eq.~(\ref{eq.reduced}), and, subsequently, Eq.~(\ref{eq.npartm}):
\begin{align}
&\quad\mathrm{Tr}^{(N)}_{N-M}\left\{\left[H^{(2)},\rho^{(N)}(t)\right]\right\}\notag\\
&=\frac{1}{M!}\sum_{\substack{n_1 \dots n_M\\m_1 \dots m_M\\k_1 \dots k_N}}\left[\sum_{n_{M+1}\dots n_N}\langle\varphi_{n_1}\dots\varphi_{n_N}|H^{(2)}|\varphi_{k_1}\dots\varphi_{k_N}\rangle\rho^{(N)}_{k_1\dots k_N;m_1\dots m_Mn_{M+1}\dots n_N}(t)\right.\notag\\
&\hspace{2cm}\left.-\sum_{m_{M+1}\dots m_N}\rho^{(N)}_{n_1\dots n_Mm_{M+1}\dots m_N;k_1\dots k_N}(t)\langle\varphi_{k_1}\dots\varphi_{k_N}|H^{(2)}|\varphi_{m_1}\dots\varphi_{m_N}\rangle\right]\notag\\&\hspace{2cm}\times a^{\dagger}_{n_1}\dots a^{\dagger}_{n_M}a_{m_1}\dots a_{m_M}\notag\\
&=\frac{1}{M!}\sum_{\substack{n_1 \dots n_M\\m_1 \dots m_M\\k_1 \dots k_N}}\notag\\
&\quad\left[\sum_{n_{M+1}\dots n_N}\sum_{1\leq\alpha_1<\alpha_2\leq N}H_{n_{\alpha_1}n_{\alpha_2};k_1k_2}^{(2)}\rho^{(N)}_{k_1k_2\{n_1\dots n_N\}\backslash\{n_{\alpha_1}n_{\alpha_2}\};m_1\dots m_Mn_{M+1}\dots n_N}(t)\right.\notag\\
&\quad\left.-\sum_{m_{M+1}\dots m_N}\sum_{1\leq\beta_1<\beta_2\leq N}\rho^{(N)}_{n_1\dots n_Mm_{M+1}\dots m_N;k_1k_2\{m_1\dots m_N\}\backslash\{m_{\beta_1}m_{\beta_2}\}}(t)H_{k_1k_2;m_{\beta_1}m_{\beta_2}}^{(2)}\right]\notag\\&\hspace{2cm}\times a^{\dagger}_{n_1}\dots a^{\dagger}_{n_M}a_{m_1}\dots a_{m_M}.
\end{align}
Again, we need to sort out which of the indices are contracted. In order to distinguish the possibilities for the different choices of $\alpha_1$, $\alpha_2$, as well as of $\beta_1$ and $\beta_2$, we note the following statement, which follows from elementary combinatorics: For $1\leq\alpha_1<\alpha_2\leq N$ and $M\leq N$, the following holds:
\begin{enumerate}
\item There are $\binom{M}{2}$ possibilities to find both $\alpha_1$ and $\alpha_2\leq M$.
\item There are $M(N-M)$ possibilities to find $\alpha_1\leq M$ and $\alpha_2> M$.
\item There are $\binom{N-M}{2}$ possibilities to find both $\alpha_1$ and $\alpha_2>M$.
\end{enumerate}
This yields
\begin{align}\label{eq.initialstatecondition}
&\mathrm{Tr}^{(N)}_{N-M}\left\{\left[H^{(2)},\rho^{(N)}(t)\right]\right\}\\
&=\frac{1}{M!}\sum_{\substack{n_1 \dots n_M\\m_1 \dots m_M\\k_1 k_2}}\left[\binom{M}{2}H_{n_{1}n_{2};k_1k_2}^{(2)}\rho^{(M)}_{k_1k_2n_3\dots n_N;m_1\dots m_M}(t)\right.\notag\\
&\hspace{1cm}+M(N-M)\sum_{n_{M+1}}H_{n_{1}n_{M+1};k_1k_2}^{(2)}\rho^{(M+1)}_{k_1k_2n_2\dots n_M;m_1\dots m_Mn_{M+1}}(t)\notag\\
&\hspace{1cm}+\binom{N-M}{2}\sum_{n_{M+1}}\sum_{n_{M+2}}H_{n_{M+1}n_{M+2};k_1k_2}^{(2)}\rho^{(M+2)}_{k_1k_2n_1\dots n_M;m_1\dots m_Mn_{M+1}n_{M+2}}(t)\notag\\
&\hspace{1cm}-\binom{M}{2}\rho^{(M)}_{n_1\dots n_M;k_1k_2m_3\dots m_M}(t)H_{k_1k_2;m_{1}m_{2}}^{(2)}\notag\\
&\hspace{1cm}-M(N-M)\sum_{m_{M+1}}\rho^{(M+1)}_{n_1\dots n_Mm_{M+1};k_1k_2m_2\dots m_M}(t)H_{k_1k_2;m_{1}m_{M+1}}^{(2)}\notag\\
&\hspace{1cm}\left.-\binom{N-M}{2}\sum_{m_{M+1}}\sum_{m_{M+2}}\rho^{(M+2)}_{n_1\dots n_Mm_{M+1} m_{M+2};k_1k_2m_1\dots m_M}(t)H_{k_1k_2;m_{M+1}m_{M+2}}^{(2)}
\right]\notag\\&\hspace{2cm}\times a^{\dagger}_{n_1}\dots a^{\dagger}_{n_M}a_{m_1}\dots a_{m_M}\notag\\
&=\left[H^{(2)},\rho^{(M)}(t)\right]\notag\\
&\quad+\frac{M(N-M)}{M!}\sum_{\substack{n_1,\dots,n_M\\m_1,\dots,m_M\\k_1,k_2}}\left[\sum_{n_{M+1}}H_{n_{1}n_{M+1};k_1k_2}^{(2)}\rho^{(M+1)}_{k_1k_2n_2\dots n_M;m_1\dots m_Mn_{M+1}}(t)\right.\notag\\
&\quad-\left.\sum_{m_{M+1}}\rho^{(M+1)}_{n_1\dots n_Mm_{M+1};k_1k_2m_2\dots m_M}(t)H_{k_1k_2;m_{1}m_{M+1}}^{(2)}\right]a^{\dagger}_{n_1}\dots a^{\dagger}_{n_M}a_{m_1}\dots a_{m_M}\notag.
\end{align}
Terms containing the $M$-particle reduced density matrix can again be identified with the commutator $[H^{(2)},\rho^{(M)}(t)]$. This involves the representation of the two-particle Hamiltonian $H^{(2)}$ in an $M$-particle basis, giving rise to the prefactors $\binom{M}{2}$, in agreement with Eqs.~(\ref{eq.ms1}) and (\ref{eq.ms2}). Notice that the contributions containing $\rho^{(M+2)}$ cancel.

We abbreviate the second term [the last two lines in the expression~(\ref{eq.initialstatecondition}) above] by introducing the symbol $\mathfrak{L}^{(M)}(H^{(2)},\rho^{(M+1)}(t))$, leading to
\begin{align}\label{eq.mplusoneterm}
\mathrm{Tr}^{(N)}_{N-M}\left\{\left[H^{(2)},\rho^{(N)}(t)\right]\right\}=\left[H^{(2)},\rho^{(M)}(t)\right]+\frac{M(N-M)}{M!}\mathfrak{L}^{(M)}(H^{(2)},\rho^{(M+1)}(t)).
\end{align}
The right-hand side contains, beyond the commutator involving the $M$-particle density matrix, also contributions of the $(M+1)$-particle density matrix. To find a compact expression for $\mathfrak{L}^{(M)}(H^{(2)},\rho^{(M+1)}(t))$, we consider the special case $N=M+1$, with $M\geq 2$ in Eq.~(\ref{eq.mplusoneterm}), which yields
\begin{align}
\mathfrak{L}^{(M)}(H^{(2)},\rho^{(M+1)}(t))=\frac{M!}{M}\left(\mathrm{Tr}^{(M+1)}_1\left\{\left[H^{(2)},\rho^{(M+1)}(t)\right]\right\}-\left[H^{(2)},\rho^{(M)}(t)\right]\right)
\end{align}
Inserting this back into Eq.~(\ref{eq.mplusoneterm}) for arbitrary $N$, leads to
\begin{align}\label{eq.twopartcontr}
&\quad\mathrm{Tr}^{(N)}_{N-M}\left\{\left[H^{(2)},\rho^{(N)}(t)\right]\right\}\notag\\
&=\left[H^{(2)},\rho^{(M)}(t)\right]+(N-M)\left(\mathrm{Tr}^{(M+1)}_1\left\{\left[H^{(2)},\rho^{(M+1)}(t)\right]\right\}-\left[H^{(2)},\rho^{(M)}(t)\right]\right)\notag\\
&=\left[(1-(N-M))H^{(2)},\rho^{(M)}(t)\right]+(N-M)\mathrm{Tr}^{(M+1)}_1\left\{\left[H^{(2)},\rho^{(M+1)}(t)\right]\right\}.
\end{align}
Together with the single-particle term~(\ref{eq.singleparticlebbgky}), this transforms~(\ref{eq.startingequationbbgky}) into a hierarchical expansion of the reduced dynamics,
\begin{align}\label{eq.bbgky}
i\hbar\frac{\partial}{\partial t}\rho^{(M)}(t)&=\left[H^{(1)}+(1-(N-M))H^{(2)},\rho^{(M)}(t)\right]\notag\\&\quad+(N-M)\mathrm{Tr}^{(M+1)}_1\left\{\left[H^{(2)},\rho^{(M+1)}(t)\right]\right\},
\end{align}
which depends on $\rho^{(M)}$ and $\rho^{(M+1)}$. A hierarchal expansion of this type was first derived to describe the statistical physics of classical systems \cite{Yvon1935,Bogoliubov1946,Born1946,Kirkwood1946} and is commonly referred to as the BBGKY hierarchy. It was later generalized to quantum systems \cite{Bogoliubov1947}; see also \cite{Bogoliubov,Roepke2013}. 

Note that, whenever the reduced single-particle dynamics is considered ($M=1$), the two-particle Hamiltonian part drops out in Eq.~(\ref{eq.bbgky}), since both sides of the equation are projected onto the single-particle subspace. We obtain for $M=1$ 
\begin{align}
i\hbar\frac{\partial}{\partial t}\rho^{(1)}(t)=\left[H^{(1)},\rho^{(1)}(t)\right]+(N-1)\mathrm{Tr}^{(2)}_1\left\{\left[H^{(2)},\rho^{(2)}(t)\right]\right\}
\end{align}
Moreover, if no particle is traced out, that is when $N=M$, the evolution~(\ref{eq.bbgky}) reduces to the unperturbed von Neumann equation~(\ref{eq.vnspte}) for the $N$-particle density operator.

\subsection{Bosonic product states}
To obtain a closed expression for the dynamics of $\rho^{(M)}(t)$, the higher-order terms, i.e., those that depend on $M+1$ particles, have to be expressed as a function of $\rho^{(M)}(t)$. This requires an approximation that truncates this hierarchy~\cite{Bogoliubov,Bogoliubov2,Roepke2013}. One possibility to achieve this, is to assume a mean-field-type description of the system, in which all particles are described by the same single-particle state. The quantum statistics of bosonic particles, however, sets significant limitations on the set of possible many-body quantum states. This problem is related to the more general question: Which $M$-particle states can be obtained by performing the partial trace over $N-M$ particles of an $N$-particle density operator \cite{RevModPhys.35.668,Garrod,Kummer1967,Kummer1970,DavidsonBook,Radzki2010,RDMBook}?

The complexity of this well-known problem is entirely generated by the particle statistics. When dealing with distinguishable particles, we can readily express any uncorrelated $N$-particle state by the tensor product of $N$ single-particle density operators. For identical particles, the set of quantum states is limited since they must obey the \mbox{(anti-)}symmetrization rules. In the case of bosonic particles, all quantum states must be constructed as convex linear combinations of the symmetrized states~(\ref{eq.creatvacsym}). Note that even a seemingly symmetric construction, such as $\rho_p^{(2)}=\rho^{(1)}\otimes\rho^{(1)}$, can in general not be represented in terms of symmetrized states, and is therefore not a valid bosonic product state. The reason is that the symmetrization rule must not be applied on the level of operators (i.e. it does not involve permutations of the single-particle density matrices), but instead on the level of states. The non-equivalence of the two is easily seen when each of the single-particle density matrices is expressed in terms of pure states by means of their spectral decomposition. 

Another way to see that $\rho_p^{(2)}=\rho^{(1)}\otimes\rho^{(1)}$ does not yield a proper bosonic product state is to perform the partial trace operation~(\ref{eq.reduced}), which leads to the single-particle state
\begin{align}\label{eq.redstated}
\mathrm{Tr}^{(2)}_1\{\rho^{(2)}_p\}=\frac{1}{2}\left(\rho^{(1)}\mathrm{Tr}^{(1)}\rho^{(1)}+(\rho^{(1)})^2\right),
\end{align}
Assuming proper normalization of the single-particle density matrix, $\mathrm{Tr}^{(1)}\rho^{(1)}=1$, the above result implies that we recover the single-particle density operator $\rho^{(1)}$ \textit{only} when $(\rho^{(1)})^2=\rho^{(1)}$, i.e., when $\rho^{(1)}$ is a projector describing a pure state $|\Phi\rangle$. In fact, only in this case does $\rho_p^{(2)}$ respect the symmetrization rules for bosonic operators. The result~(\ref{eq.redstated}) can be generalized to (anti-)symmetric product states of an arbitrary number $N$ of bosonic or fermionic particles in terms of a recurrence formula, see Theorem 3.1 in~\cite{Radzki2010}.

In this specific case of a pure state, the elements of the single-particle density matrix can be written as $\rho^{(1)}_{ij}=\langle\varphi_i|\Phi\rangle\langle\Phi|\varphi_j\rangle$. Introducing $a^{\dagger}_{\Phi}|0\rangle=|\Phi\rangle$, we obtain $\rho^{(1)}=a^{\dagger}_{\Phi}a_{\Phi}$, which we can use to define a general \textit{bosonic pure product state} as
\begin{align}\label{eq.pureproductstate}
\rho^{(N)}_P=\frac{1}{N!}:(\rho^{(1)})^N:=\frac{1}{N!}\underbrace{a^{\dagger}_{\Phi}\dots a^{\dagger}_{\Phi}}_{\text{N times}}\underbrace{a_{\Phi}\dots a_{\Phi}}_{\text{N times}},
\end{align}
where the double dots denote normal ordering. The many-body matrix elements in a basis of symmetrized states read
\begin{align}\label{eq.matrixelspps}
\langle\varphi_{i_1}\dots\varphi_{i_N}|\rho^{(N)}_P|\varphi_{j_1}\dots\varphi_{j_N}\rangle=c_{i_1}\cdots c_{i_N}c^*_{j_1}\cdots c^*_{j_N},
\end{align}
with $c_i=\langle\varphi_i|\Phi\rangle$. The many-body system is described by such a pure product state when a particular single-particle state is macroscopically occupied, i.e., when the system describes a perfect Bose-Einstein condensate. This can be interpreted as a mean-field ansatz, where all individual particles are described by the same quantum state. 

The bosonic pure product state~(\ref{eq.pureproductstate}) represents the only quantum state which defines the entire symmetric many-body quantum state uniquely on the basis of the single-particle state $\rho^{(1)}$ while at the same time its single-particle distribution also coincides with $\rho^{(1)}$. For fermionic systems, the anti-symmetrization rules exclude any similar construction. The combination of these two properties is important for the efficient microscopic description of the subdynamics. A similar issue arises in the standard theory of open quantum systems, where a closed description of the open-system subdynamics in terms of reduced density operators of the open system requires statistical independence between system and environment \cite{Lindblad1996}.

Based on the pure product state ansatz, we can decompose the elements of the $N$-particle density matrix $\rho^{(N)}$ in terms of the single-particle entries according to Eq.~(\ref{eq.matrixelspps}). On the one hand, this allows us to find a closed description for the subdynamics of $\rho^{(M)}$. Being bound to the pure state assumption for the single-particle state at all times may be adequate for coherent mean-field descriptions, to be discussed in the following section. On the other hand, this limits our possibilities to extend the present approach towards an effective description of incoherent subdynamics, which we will discuss in section~\ref{sec.identicalparticlesmeq}.

\subsection{Coherent mean-field evolution}
Employing the pure product state assumption~(\ref{eq.pureproductstate}), we truncate the hierarchical family of dynamical equations~(\ref{eq.bbgky}). While, in principle, we could separate the full density matrix into contributions of single-particle pure state contributions according to Eq.~(\ref{eq.matrixelspps}), we use this decomposition to separate $\rho^{(M+1)}$ into the contributions of $\rho^{(1)}$ and $\rho^{(M)}$, and retain the density matrix notation, keeping in mind that the state is pure. The reduced dynamics can then be written as
\begin{align}\label{eq.truncatedbbgky}
i\hbar\frac{\partial}{\partial t}\rho^{(M)}(t)=\left[H^{(1)}+H^{(2)}+(N-M)C^{(1)}(t),\rho^{(M)}(t)\right],
\end{align}
where $C^{(1)}(t)$ is defined as
\begin{align}\label{eq.copsp}
C^{(1)}(t)=\sum_{ijmn}H^{(2)}_{nj;im}\rho^{(1)}_{ij}(t)a^{\dagger}_na_m.
\end{align}
Equation~(\ref{eq.truncatedbbgky}) still has the form of the von Neumann equation. However, the mean-field potential term $C^{(1)}(t)$, which itself depends on $\rho^{(1)}$, is added to the unperturbed Hamiltonian. For $M=1$, we insert a pure state,
\begin{align}\label{eq.puresingleparticlestate}
\rho^{(1)}(t)=|\Phi(t)\rangle\langle\Phi(t)|,
\end{align}
into Eq.~(\ref{eq.truncatedbbgky}). Then, the evolution is equivalent to the one predicted by the following \textit{nonlinear mean-field Schr\"odinger equation}, 
\begin{align}\label{eq.nlseq}
i\hbar\frac{\partial}{\partial t}|\Phi(t)\rangle=H^{(1)}_{\mathrm{eff}}(t)|\Phi(t)\rangle,
\end{align}
with the effective single-particle Hamiltonian,
\begin{align}\label{eq.Heff}
H^{(1)}_{\mathrm{eff}}(t)=H^{(1)}+(N-1)C^{(1)}(t),
\end{align}
where $C^{(1)}(t)$ is a function of $|\Phi(t)\rangle$, and hence contributes a nonlinear effective potential that accounts for the particle-particle interactions.

For completeness, we note that a similar calculation in the interaction picture~(\ref{eq.vNinter}) yields
\begin{align}\label{eq.bbgkyintpicpps}
i\hbar\frac{\partial}{\partial t}\rho_I^{(M)}(t)&=\mathrm{Tr}^{(N)}_{N-M}\left\{\left[H_I^{(2)}(t),\rho_I^{(N)}(t)\right]\right\}\notag\\&=\left[H_I^{(2)}(t)+(N-M)C_I^{(1)}(t,t),\rho_I^{(M)}(t)\right],
\end{align}
where the nonlinear potential in the interaction picture can be expressed as
\begin{align}\label{eq.cop}
C_I^{(1)}(t_1,t_2)=\sum_{ijnm}H^{(2)}_{I,nj;im}(t_1)\rho^{(1)}_{I,ij}(t_2)a^{\dagger}_na_m,
\end{align}
and we denote the matrix elements of the corresponding interaction-picture operators as $\rho^{(1)}_{I,ij}(t)=\langle\varphi_i|\rho^{(1)}_I(t)|\varphi_j\rangle$ and $H^{(2)}_{I,nj;im}(t)=\langle\varphi_n\varphi_j|H^{(2)}_I(t)|\varphi_i\varphi_m\rangle$.

The dynamical equation~(\ref{eq.bbgky}) describes the microscopically derived, completely general quantum dynamics of bosonic subsystems, without assumptions on the potential landscape or the nature of the pairwise interaction. The obtained BBGKY-type hierarchy can be truncated by the pure product state assumption~(\ref{eq.pureproductstate}). This yields the nonlinear mean-field Schr\"odinger equation~(\ref{eq.nlseq}), which governs the dynamics of a macroscopically occupied single-particle state, for arbitrary many-body Hamiltonians~(\ref{eq.generalmanybodyH}) of interacting bosons. The emergent nonlinear term $C^{(1)}$ is reminiscent of the Gross-Pitaevskii equation, and, in fact, we will show in the next section that the general nonlinear mean-field Schr\"odinger equation~(\ref{eq.nlseq}) reduces to the Gross-Pitaevskii equation in the special case of a contact interaction potential.

\subsection{Special case: The Gross-Pitaevskii equation}\label{sec.frombbgkytogpe}
Let us consider the special case of a contact interaction $v(\mathbf{x}_1,\mathbf{x}_2)=g\delta(\mathbf{x}_2-\mathbf{x}_1)$, with constant $g$ \cite{PitS03}. We choose a representation in terms of field operators $\hat{\Psi}(\mathbf{x})=\sum_i\varphi_i(\mathbf{x})a_i$, where $\varphi_i(\mathbf{x})=\langle \mathbf{x}|\varphi_i\rangle$, and $|\mathbf{x}\rangle$ denotes an eigenstate of the position operator. The interaction Hamiltonian then reads
\begin{align}
H^{(2)}&=\frac{1}{2}\int d\mathbf{x}_1\int d\mathbf{x}_2\hat{\Psi}^{\dagger}(\mathbf{x}_1)\hat{\Psi}^{\dagger}(\mathbf{x}_2)v(\mathbf{x}_1,\mathbf{x}_2)\hat{\Psi}(\mathbf{x}_1)\hat{\Psi}(\mathbf{x}_2)\notag\\&=\frac{g}{2}\int d\mathbf{x}\hat{\Psi}^{\dagger}(\mathbf{x})\hat{\Psi}^{\dagger}(\mathbf{x})\hat{\Psi}(\mathbf{x})\hat{\Psi}(\mathbf{x}).
\end{align}
Inserting this into the effective mean-field term~(\ref{eq.copsp}) yields
\begin{align}\label{eq.copcontactint}
&\quad C^{(1)}(t)\notag\\&=g\sum_{ijnm}\int d\mathbf{x}\int d\mathbf{x}_1\int d\mathbf{x}_2\varphi^*_n(\mathbf{x})\varphi^*_j(\mathbf{x})\varphi_i(\mathbf{x})\varphi_m(\mathbf{x})\varphi^*_i(\mathbf{x}_1)\rho^{(1)}(t;\mathbf{x}_1;\mathbf{x}_2)\varphi_j(\mathbf{x}_2) a^{\dagger}_na_m\notag\\
&=g\int d\mathbf{x} \hat{\Psi}^{\dagger}(\mathbf{x})\hat{\Psi}(\mathbf{x})|\Phi(t,\mathbf{x})|^2,
\end{align}
where we used the completeness relation~(\ref{eq.completenessrelation}) and Eq.~(\ref{eq.puresingleparticlestate}). The nonlinear mean-field Schr\"odinger equation~(\ref{eq.nlseq}) predicts the following time evolution of the single-particle wavefunction:
\begin{align}\label{eq.derivedgpemean}
i\hbar\frac{\partial}{\partial t}\Phi(t,\mathbf{x})&=\langle\mathbf{x}|H^{(1)}_{\mathrm{eff}}(t)|\Phi(t)\rangle\notag\\
&=\left[-\frac{\hbar^2}{2m}\nabla_{\mathbf{x}}^2+V_0(\mathbf{x})+g(N-1)|\Phi(t,\mathbf{x})|^2\right]\Phi(t,\mathbf{x}).
\end{align}
Here we have explicitly written out the kinetic terms $-(\hbar^2/2m)\nabla_{\mathbf{x}}^2$ and the single-particle potential $V_0(\mathbf{x})$ contained in $H^{(1)}$. Recall that by virtue of the pure product state assumption~(\ref{eq.pureproductstate}), $\Phi(t,\mathbf{x})=\langle\mathbf{x}|\Phi(t)\rangle$ describes an $N$-fold occupied single-particle state. When $N\gg 1$, this occupation becomes macroscopic and Eq.~(\ref{eq.derivedgpemean}) coincides with the Gross-Pitaevskii equation \cite{Gross1961,Pitaevskii1961,PitS03}. Unlike standard approaches \cite{PitS03}, the above derivation does not involve the controversial \cite{RevModPhys.73.307} assumption of a coherent superposition between states of different particle numbers---for a recent discussion see \cite{Gertjerenken2015}.

\section{Second-order master equation for interacting bosons: Towards the description of incoherent effects}\label{sec.identicalparticlesmeq}
The partial trace operation~(\ref{eq.reduced}) over the many-body von Neumann equation~(\ref{eq.vnspte}), together with the pure product state ansatz~(\ref{eq.pureproductstate}) allowed us to derive a nonlinear mean-field equation~(\ref{eq.truncatedbbgky}) whose evolution is fully coherent at all times. To extend our treatment to account for interaction-induced incoherent effects on the reduced state level, we now perform a second-order perturbative expansion in terms of the interaction Hamiltonian $H^{(2)}$ in Eq.~(\ref{eq.generalmanybodyH}), before performing the partial trace operation. This is the approach pursued in microscopic derivations\cite{BreuerPetruccione2006} of Lindblad-type master equations \cite{Lindblad1976,GKS1976} for the description of open quantum systems where the system degrees of freedom are distinct from those of the environment. In such master equations the coherent dynamics in the system degrees of freedom is amended by incoherent terms with the characteristic Lindblad structure.

\subsection{$M$-particle subdynamics of an $N$-particle bosonic system}
A convenient point of departure to develop a second-order perturbative expansion in $H^{(2)}$ is the interaction-picture. We expand expression~(\ref{eq.vNinter}) to second order, before performing the partial trace~(\ref{eq.reduced}) on both sides of the equation:
\begin{align}
\frac{\partial}{\partial t}\rho_I^{(M)}(t)&=\mathrm{Tr}^{(N)}_{N-M}\left\{-\frac{i}{\hbar}\left[H_I^{(2)}(t),\rho_I^{(N)}(0)\right]\right\}-\frac{1}{\hbar^2}\int_0^tds\mathrm{Tr}^{(N)}_{N-M}[H_I^{(2)}(t),[H_I^{(2)}(s),\rho_I^{(N)}(s)]]\label{eq.secondorderexpression}.
\end{align}
Before truncating the perturbative expansion by replacing the time argument of $\rho_I^{(N)}(s)$ in the integral with $t$, this is an exact expression in the considered subspace of $M$ particles. The term containing the initial condition, $\mathrm{Tr}^{(N)}_{N-M}\{[H_I^{(2)}(t),\rho_I^{(N)}(0)]\}$, has the same structure as those that were discussed in Eq.~(\ref{eq.twopartcontr}). The remaining two terms which emerge from the double commutator are due to the perturbative second-order expansion and produce incoherent terms with Lindblad structure, aside from coherent commutator contributions. The explicit determination of these two terms is a rather laborious task. Yet, it requires only elementary methods, akin to those already employed in the derivation of Eq.~(\ref{eq.bbgky}). A detailed presentation of the intermediate steps can be found in \cite{GessnerDiss}. The result is again of hierarchical type. While we already noted in Eqs.~(\ref{eq.initialstatecondition}) and (\ref{eq.bbgky}) that a first-order expression (i.e., a single commutator with $H^{(2)}$) generates terms that contain the reduced density operators of $M+1$ and $M+2$ particles, the highest order, i.e., $M+2$, eventually vanishes. In the second-order expression (i.e., the double commutator with $H^{(2)}$) contributions of up to $M+4$ particles can be generated, but, again, the highest order vanishes. Hence, in the second-order expression we encounter contributions of up to $M+3$ particles. To be able to find a closed expression for the subdynamics of $\rho^{(M)}$, we need to express these higher-order contributions as a function of $\rho^{(M)}$. This is once again achieved by means of the pure product state assumption~(\ref{eq.pureproductstate}). We finally obtain:
\begin{align}
\quad\frac{\partial}{\partial t}\rho_I^{(M)}(t)&=-\frac{i}{\hbar}\left[H_I^{(2)}(t),\rho_I^{(M)}(0)\right]-\frac{1}{\hbar^2}\int_0^{t}ds\left[H_I^{(2)}(t),\left[H_I^{(2)}(s),\rho_I^{(M)}(s)\right]\right]\label{eq.rhomdot}\\
&\quad-(N-M)\frac{i}{\hbar}[C_I^{(1)}(t,0),\rho_I^{(M)}(0)]\notag\\
&\quad-(N-M)\frac{1}{\hbar^2}\int_0^tds\sum_{\alpha\beta}[B^{(1)}_{[\beta\alpha]}(t),[A^{(1)}_{[\alpha\beta]}(s,s),\rho_I^{(M)}(s)]]\notag\\
&\quad-(N-M)\frac{1}{\hbar^2}\int_0^tds[H_I^{(2)}(t),[C_I^{(1)}(s,s),\rho_I^{(M)}(s)]]\notag\\
&\quad-(N-M)\frac{1}{\hbar^2}\int_0^tds[C_I^{(1)}(t,s),[H_I^{(2)}(s),\rho_I^{(M)}(s)]]\notag\\
&\quad-(N-M)(N-M-1)\frac{1}{\hbar^2}\int_0^tds[C_I^{(1)}(t,s),[C_I^{(1)}(s,s),\rho_I^{(M)}(s)]]\notag\\
&\quad-(N-M)(N-M-1)\frac{i}{\hbar}\int_0^tds[H_D^{(1)}(t,s),\rho_I^{(M)}(s)]\notag\\
&\quad-(N-M)(N-M-1)\frac{i}{\hbar}\int_0^tds[H_S^{(2)}(t,s),\rho_I^{(M)}(s)]\notag\\
&\quad-(N-M)(N-M-1)(N-M-2)\frac{i}{\hbar}\int_0^tds[H_E^{(1)}(t,s),\rho_I^{(M)}(s)],\notag
\end{align}
where we introduced the single-particle operators
\begin{align}\label{eq.boperators}
B^{(1)}_{[\beta\alpha]}(t)=\sum_{ij}H^{(2)}_{I,i\beta; j\alpha}(t)a^{\dagger}_ia_j.
\end{align}
and
\begin{align}\label{eq.aoperators}
A^{(1)}_{[\alpha\beta]}(s,t):=\sum_{ijk}H_{I,i\alpha; jk}^{(2)}(s)\rho^{(1)}_{I,k\beta}(t)a^{\dagger}_ia_j=\sum_kB^{(1)}_{[\alpha k]}(s)\rho_{I,k\beta}^{(1)}(t).
\end{align}
The indices are written in angular brackets to emphasize that the $A^{(1)}_{[\alpha\beta]}$ and $B^{(1)}_{[\beta\alpha]}$ are not matrix elements of an operator, but full single-particle operators. The operators $B^{(1)}_{[\alpha\beta]}$ exhibit the properties $B^{(1)\dagger}_{[\alpha\beta]}=B^{(1)}_{[\beta\alpha]}$ and $\sum_{\alpha}B^{(1)}_{[\alpha\alpha]}(t)=\mathrm{Tr}^{(2)}_1H^{(2)}_I(t)$. Moreover, we encounter additional Hermitian operators
\begin{align}
H^{(1)}_D(t_1,t_2)&=-\frac{i}{\hbar}(D^{(1)}(t_1,t_2)-D^{(1)}(t_1,t_2)^{\dagger})\notag\\&=\sum_{ij}\mathrm{Tr}^{(1)}\{B^{(1)}_{[ij]}(t_1)\mathrm{Tr}_1^{(2)}\left\{-\frac{i}{\hbar}[H_I^{(2)}(t_2),\rho_I^{(2)}(t_2)]\right\}\}a^{\dagger}_ia_j,\label{eq.defhd}\\
H_S^{(2)}(t_1,t_2)&=-\frac{i}{\hbar}\left(S^{(2)}(t_1,t_2)-S^{(2)}(t_1,t_2)^{\dagger}\right),\label{eq.defh2s}
\intertext{and}
H_E^{(1)}(t_1,t_2)&=-\frac{i}{\hbar}(E^{(1)}(t_1,t_2)-E^{(1)}(t_1,t_2)^{\dagger})\label{eq.defHE},
\end{align}
defined via
\begin{align}
D^{(1)}(t_1,t_2)&=\sum_{ij}\mathrm{Tr}^{(1)}\{B^{(1)}_{[ij]}(t_1)\mathrm{Tr}_1^{(2)}\{H_I^{(2)}(t_2)\rho_I^{(2)}(t_2)\}\}a^{\dagger}_ia_j,\label{eq.doperator}\\
E^{(1)}(t_1,t_2)&=\frac{1}{2}C_I^{(1)}(t_2,t_2)\rho_I^{(1)}(t_2)C_I^{(1)}(t_1,t_2),\label{eq.eoperator}\\
S^{(1)}(t_1,t_2)&=\frac{1}{2}H_I^{(2)}(t_1)\rho_I^{(2)}(t_2)H_I^{(2)}(t_2),\label{eq.soperator}
\end{align}
and $C^{(1)}_I(t_1,t_2)$ was already introduced in Eq.~(\ref{eq.cop}). 

Within the $M$-particle subspace, the expression~(\ref{eq.rhomdot}) is exact, except for the pure product state assumption~(\ref{eq.pureproductstate}). On top of the nonlinear amendments to the Hamiltonian, such as $C_I^{(1)}$, which we encountered already in the coherent mean-field equation~(\ref{eq.bbgkyintpicpps}), the result~(\ref{eq.rhomdot}) contains also double-commutator operations on $\rho^{(M)}$. These are known as `Lindblad dissipators' from standard open-system theory\cite{BreuerPetruccione2006} and lead to incoherent, irreversible dynamics. As before, due to the factorization~(\ref{eq.matrixelspps}) the matrix elements of $\rho^{(M)}$ could be further decomposed into the individual pure-state contributions of single particles.

If no particle is traced over, i.e., if $N=M$, only the first line of Eq.~(\ref{eq.rhomdot}) remains. Indeed, these terms describe the fully coherent dynamics of the $N$-particle density matrix according to the von Neumann equation. The remaining terms can be classified depending on the number of additional particles which need to be taken into account. For example, if only one additional particle is present in the system, i.e., if $N=M+1$, the last four lines of Eq.~(\ref{eq.rhomdot}) vanish and only those terms proportional to the prefactor $N-M$ remain. Each of these terms indeed contains the contribution of exactly one additional particle. This is manifested by the dependence of $A^{(1)}_{[\alpha\beta]}$ [Eq.~(\ref{eq.aoperators})] and $C^{(1)}_I$ [Eq.~(\ref{eq.cop})] on $\rho_I^{(1)}$. These terms, moreover, exhibit the anticipated double-commutator structure, which is equivalent to a master equation with Lindblad structure, as we will discuss later in further detail. The next three terms contain the contribution of two additional particles, via two occurrences of $C^{(1)}$ and the respective definitions Eq.~(\ref{eq.doperator}) and~(\ref{eq.soperator}). Finally, the last term depends on the state of three additional particles, see Eq.~(\ref{eq.eoperator}). The dependence on higher particle numbers generate nonlinearities, as already observed in previous sections.

\subsection{Dissipative mean-field master equation for dilute gases}
Elements of the non-perturbative mean-field equation~(\ref{eq.bbgkyintpicpps}) can be identified also in the second-order master equation~(\ref{eq.rhomdot}), by regrouping terms as follows:
\begin{align}\label{eq.mostgeneralproductstateevolution_simpler}
\frac{\partial}{\partial t}\rho_I^{(M)}(t)&=-\frac{i}{\hbar}\left[H_I^{(2)}(t)+(N-M)C_I^{(1)}(t,0),\rho_I^{(M)}(0)\right]\\
&\quad-\frac{1}{\hbar^2}\int_0^{t}ds\left[H_I^{(2)}(t),\left[H_I^{(2)}(s)+(N-M)C_I^{(1)}(s,s),\rho_I^{(M)}(s)\right]\right]\notag\\
&\quad-\frac{1}{\hbar^2}\int_0^tds\left[(N-M)C_I^{(1)}(t,s),\left[H_I^{(2)}(s)+(N-M-1)C_I^{(1)}(s,s),\rho_I^{(M)}(s)\right]\right]\notag\\
&\quad-(N-M)\frac{1}{\hbar^2}\int_0^tds\sum_{\alpha\beta}[B^{(1)}_{[\beta\alpha]}(t),[A^{(1)}_{[\alpha\beta]}(s,s),\rho_I^{(M)}(s)]]\notag\\
&\quad-(N-M)(N-M-1)\frac{i}{\hbar}\int_0^tds[H_D^{(1)}(t,s),\rho_I^{(M)}(s)]\notag\\
&\quad-(N-M)(N-M-1)\frac{i}{\hbar}\int_0^tds[H_S^{(2)}(t,s),\rho_I^{(M)}(s)]\notag\\
&\quad-(N-M)(N-M-1)(N-M-2)\frac{i}{\hbar}\int_0^tds[H_E^{(1)}(t,s),\rho_I^{(M)}(s)].\notag
\end{align}
In order to approximate this expression we consider the following assumptions:
\begin{itemize}
\item $N-M\gg 1$, i.e., the total number of particles is large compared to the observable subspace. This allows us to replace $N-M-1\approx N-M$.
\item $C_I^{(1)}(t,s) \rightarrow C_I^{(1)}(t,t)$, and $C_I^{(1)}(t,0) \rightarrow C_I^{(1)}(t,t)$. Recall from Eq.~(\ref{eq.cop}) that the first, i.e., left time argument of $C_I^{(1)}$ refers to the interaction-picture Hamiltonian $H_I^{(2)}$, while the second time argument relates to the quantum state $\rho^{(1)}_I$. Thus, the above replacements effectively produce a time-dependent effective mean-field potential, which is generated by the contributions of the quantum state $\rho_I^{(1)}$ on the right-hand side.
\end{itemize}
Using these approximations, we can rewrite the first three lines of Eq.~(\ref{eq.mostgeneralproductstateevolution_simpler}) as
\begin{align}
&-\frac{i}{\hbar}\left[H_I^{(2)}(t)+(N-M)C_I^{(1)}(t,t),\rho_I^{(M)}(0)\right]\notag\\
&-\frac{1}{\hbar^2}\int_0^{t}ds\left[H_I^{(2)}(t),\left[H_I^{(2)}(s)+(N-M)C_I^{(1)}(s,s),\rho_I^{(M)}(s)\right]\right]\notag\\
&-\frac{1}{\hbar^2}\int_0^tds\left[(N-M)C_I^{(1)}(t,t),\left[H_I^{(2)}(s)+(N-M)C_I^{(1)}(s,s),\rho_I^{(M)}(s)\right]\right]\notag\\
=\quad&-\frac{i}{\hbar}\left[H_I^{(2)}(t)+(N-M)C_I^{(1)}(t,t),\right.\notag\\
&\hspace{1cm}\left.\rho_I^{(M)}(0)-\frac{i}{\hbar}\int_0^tds\left[H_I^{(2)}(s)+(N-M)C_I^{(1)}(s,s),\rho_I^{(M)}(s)\right]\right]\notag\\
\approx\quad&-\frac{i}{\hbar}\left[H_I^{(2)}(t)+(N-M)C_I^{(1)}(t,t),\rho_I^{(M)}(t)\right]\label{eq.integrationtrick}.
\end{align}
In the last step, we have neglected contributions of the other elements of the complete master equation~(\ref{eq.mostgeneralproductstateevolution_simpler}), i.e., effectively equating the expression~(\ref{eq.integrationtrick}) with $\partial_t\rho_I^{(M)}(t)$ to carry out the integration.

Next, following the argument presented in \cite{Bogoliubov}, we further assume a dilute sample of low particle density $\varrho=N/V\ll 1$, where $V$ denotes the particles' confining volume. The particle density function is given by $\varrho(\mathbf{x})=N\rho^{(1)}(\mathbf{x};\mathbf{x})$, where $\rho^{(1)}(\mathbf{x};\mathbf{x})$ is the position representation of the single-particle density matrix and is interpreted as the probability to find one particle at coordinate $\mathbf{x}$. Similarly, $\rho^{(S)}(\mathbf{x}_1,\dots,\mathbf{x}_S;\mathbf{x}_1,\dots,\mathbf{x}_S)$ defines the probability density function of $S$ particles. To estimate the order of magnitude of the $S$-particle density matrix, consider the simple case of a uniformly distributed density, i.e., $\rho^{(1)}(\mathbf{x};\mathbf{x})=1/V$. Assuming no correlations between the particles, one finds that the probability to find one particle at each of the coordinates $\mathbf{x}_1,\dots,\mathbf{x}_S$ is consequently given by $\rho^{(S)}(\mathbf{x}_1,\dots,\mathbf{x}_S;\mathbf{x}_1,\dots,\mathbf{x}_S)=1/V^S$. As the correlations between the particles are generally expected to be weak in dilute ensembles \cite{Bogoliubov}, we can generally estimate the $S$-particle density matrix $\rho^{(S)}$ as a quantity of the order of $1/V^{S}$. For $N\gg M$, contributions in Eq.~(\ref{eq.rhomdot}) that depend on the quantum state $\rho^{(S)}$ of $S$ additional particles always occur in combination with a prefactor $N^S$. Hence, these terms will be on the order of $\varrho^{S}$. For the description of dilute gases, we neglect such terms for $S>1$. Effectively, this eliminates the commutator terms in the final three lines of Eq.~(\ref{eq.mostgeneralproductstateevolution_simpler}): $H_D$ and $H_S$ depend directly on $\rho^{(2)}$ [see Eqs.~(\ref{eq.defhd}), (\ref{eq.doperator}), and Eqs.~(\ref{eq.defh2s}), (\ref{eq.soperator}), respectively], while $H_E$ [Eqs.~(\ref{eq.defHE}) and (\ref{eq.eoperator})] contains contributions of $\rho^{(1)}$ and of $C^{(1)}$ which, in turn, depends on $\rho^{(1)}$ as well, see Eq.~(\ref{eq.cop}).

Finally, we truncate the perturbative expansion~(\ref{eq.secondorderexpression}) to second order by changing the time arguments of the density operators on the right-hand side from $s$ to $t$. Together with Eq.~(\ref{eq.integrationtrick}), we obtain the \textit{dissipative nonlinear mean-field master-equation for dilute gases}
\begin{align}\label{eq.dissmfmeq}
\frac{\partial}{\partial t}\rho_I^{(M)}(t)=\:&-\frac{i}{\hbar}\left[H_I^{(2)}(t)+(N-M)C_I^{(1)}(t,t),\rho_I^{(M)}(t)\right]\\
&-(N-M)\frac{1}{\hbar^2}\int_0^tds\sum_{\alpha\beta}[B^{(1)}_{[\beta\alpha]}(t),[A^{(1)}_{[\alpha\beta]}(s,t),\rho_I^{(M)}(t)]]\notag,
\end{align}
where the commutator term contains the nonlinear, effective mean-field potential. This term was already encountered in the nonlinear mean-field equation~(\ref{eq.bbgkyintpicpps}). The second line represents an incoherent correction with the characteristic double-commutator structure. Aside from an initial condition [terms with time argument $0$ in Eq.~(\ref{eq.mostgeneralproductstateevolution_simpler})], this double-commutator is also the only term that prevails if we consider the simplest case of $N=2$ and $M=1$ in Eq.~(\ref{eq.mostgeneralproductstateevolution_simpler}). We can always bring this term into Lindblad structure, by rewriting
\begin{align}\label{eq.doublecommutolindblad}
&\quad-\frac{1}{\hbar^2}\int_0^tds\sum_{\alpha\beta}[B^{(1)}_{[\beta\alpha]}(t),[A^{(1)}_{[\alpha\beta]}(s,t),\rho_I^{(M)}(t)]]\notag\\
&=\frac{1}{\hbar^2}\int_0^tds\sum_{\alpha\beta}\left(A^{(1)}_{[\alpha\beta]}(s,t)\rho_I^{(M)}(t)B^{(1)}_{[\beta\alpha]}(t)-B^{(1)}_{[\beta\alpha]}(t)A^{(1)}_{[\alpha\beta]}(s,t)\rho_I^{(M)}(t)\right)+\mathrm{\mathrm{H.c.}}\notag\\
&=\frac{1}{\hbar^2}\sum_{ijkl}\Gamma_{ijkl}(t)\left(a^{\dagger}_ia_j\rho_I^{(M)}(t)a^{\dagger}_ka_l-a^{\dagger}_ka_la^{\dagger}_ia_j\rho_I^{(M)}(t)\right)+\mathrm{\mathrm{H.c.}},
\end{align}
where $\mathrm{H.c.}$ denotes the Hermitian conjugate, and we defined
\begin{align}
\Gamma_{ijkl}(t)&=\int_0^tds\sum_{\alpha\beta}\langle\varphi_i|A^{(1)}_{[\alpha\beta]}(s,t)|\varphi_j\rangle\langle\varphi_k|B^{(1)}_{[\beta\alpha]}(t)|\varphi_l\rangle\notag\\
&=\int_0^tds\sum_{\alpha\beta m}H_{i\alpha; jm}^{(2)}(s)\rho^{(1)}_{I,m\beta}(t)H^{(2)}_{k\beta ;l\alpha}(t)\notag\\
&=\int_0^tds\sum_{\alpha\beta m}H_{i\alpha; jm}^{(2)}(t-s)\rho^{(1)}_{I,m\beta}(t)H^{(2)}_{k\beta; l\alpha}(t)\notag\\
&=\int_0^tds\mathrm{Tr}^{(1)}\{B^{(1)}_{[ij]}(t-s)\rho_I^{(1)}(t)B^{(1)}_{[kl]}(t)\}.\label{eq.Gamma}
\end{align}
These complex-valued functions play the role of the bath-autocorrelation functions, known from standard treatments of open quantum systems \cite{BreuerPetruccione2006}. Since in our present situation, the bosonic particles are indistinguishable and therefore form their own bath, the quantum state $\rho^{(1)}_I$ also appears in Eq.~(\ref{eq.Gamma}), and, again, causes a nonlinearity. Introducing
\begin{align}
\Gamma_{ijkl}(t)=\frac{1}{2}\gamma_{ijkl}(t)+iS_{ijkl}(t)
\end{align}
allows us to substitute
\begin{align}
\gamma_{ijkl}(t)&=\Gamma_{ijkl}(t)+\Gamma_{lkji}^*(t),\\
S_{ijkl}(t)&=\frac{1}{2i}\left(\Gamma_{ijkl}(t)-\Gamma_{lkji}^*(t)\right).
\end{align}
This separates the expression~(\ref{eq.doublecommutolindblad}) into a coherent commutator part, and an incoherent Lindblad term
\begin{align}\label{eq.meq}
&\quad-\frac{1}{\hbar^2}\int_0^tds\sum_{\alpha\beta}[B^{(1)}_{[\beta\alpha]}(t),[A^{(1)}_{[\alpha\beta]}(s,t),\rho_I^{(M)}(t)]]\notag\\&=-\frac{i}{\hbar}\left[H^{(1)}_{LS}(t),\rho_I^{(M)}(t)\right]+\frac{1}{\hbar^2}\sum_{ijkl}\gamma_{ijkl}(t)\left(a^{\dagger}_ia_j\rho_I^{(M)}(t)a^{\dagger}_ka_l-\frac{1}{2}\left\{a^{\dagger}_ka_la^{\dagger}_ia_j,\rho_I^{(M)}(t)\right\}\right),
\end{align}
where the environment-induced Hamiltonian
\begin{align}
H^{(1)}_{LS}(t)=\frac{1}{\hbar}\sum_{ijkl}S_{ijkl}(t)a^{\dagger}_ka_la^{\dagger}_ia_j,
\end{align}
commonly referred to as Lamb-Shift Hamiltonian \cite{BreuerPetruccione2006}, causes a correction to the coherent evolution and modifies the energy levels \cite{Cohen-Tannoudji1992}.

This resulting equation~(\ref{eq.meq}) has the characteristic Lindblad structure of a time-local master equation \cite{BreuerPetruccione2006}. We note the peculiar feature that, in contrast to standard open quantum systems, the time-dependence of the autocorrelation functions $\Gamma$ is not only governed by the Hamiltonian, but also depends on the state of the system, thus rendering the resulting master equation nonlinear.

We must remark that the derivation presented in this last section is not mathematically rigorous. Some of the approximations do not commute with each other, and the result seems self-contradicting: On the one hand, Eq.~(\ref{eq.dissmfmeq}) aims to describe an incoherent dynamics, while, on the other hand, its derivation required the assumption of a pure product state at all times. Moreover, since the pure product state assumption essentially decomposes the $M$-particle state into single-particle states, it is unclear whether the derived expressions, in their present form, can lead to physical insight into the $M$-particle dynamics $(M>1)$ beyond what is already contained in the single-particle dynamics. Presently, a suitable alternative to the severely limiting pure product state assumption that achieves a closed-form description of the reduced dynamics is not known to the authors. At the moment, it remains unclear whether the developed approach is applicable to mixed states in its present form, or, otherwise, if an appropriate generalization can be achieved. The loss of probability which is expected for a pure state evolving according to Eq.~(\ref{eq.dissmfmeq}), however, might be attributed to the transfer of population into other subspaces of different particle numbers, which are not considered within our description due to a projection onto the $M$-particle subspace. This projection is also the reason why decay channels do not describe particle loss or gain (the Lindblad operators~(\ref{eq.doublecommutolindblad}) conserve the particle number since they commute with the number operator [Eq.~(\ref{eq.sumexpression}) for $M=1$]).

\section{Summary and conclusion}
In this article, we treated bosonic many-body systems in terms of symmetrized states, which account for the particles' quantum statistics within a subspace of fixed particle number. This description was employed to conveniently define suitable reduced density operators, and the corresponding partial trace operation, which determine the effective distributions of a subset of $M\leq N$ bosonic particles.

The effective dynamics of the resulting reduced $M$-particle density operator was treated in two different ways. First, we traced over the von Neumann equation of the full $N$-particle quantum system. This led to a hierarchical BBGKY-type family of equations of motion, Eq.~(\ref{eq.bbgky}). This hierarchy can be truncated by a mean-field-type pure product state assumption, under which all particles are described by the same pure single-particle state at all times. This yields the \textit{nonlinear mean-field Schr\"odinger equation}~(\ref{eq.nlseq}), describing the single-particle dynamics of an interacting Bose-Einstein condensate in an arbitrary potential landscape, and subject to arbitrary pairwise interactions~(\ref{eq.generalmanybodyH}). In the special case of a contact interaction, this was shown to reproduce the well-known and widely used Gross-Pitaevskii equation~(\ref{eq.derivedgpemean}).

Second, in an attempt to extend such treatments beyond mean-field descriptions of pure single-particle states, we employed methods from the theory of open quantum systems. The double-commutator structure of a second-order expansion in terms of the interaction Hamiltonian leads to the emergence of the characteristic Lindblad dissipators, governing the effective subdynamics. However, a closed description that avoids hierarchical families of equations, again, requires the assumption of a pure product state, which conflicts with the aspired description of interaction-induced decoherence and the implied loss of purity. Introducing further mean-field approximations, we obtained the \textit{dissipative nonlinear mean-field master-equation for dilute gases}, Eq.~(\ref{eq.dissmfmeq}), which extends the nonlinear mean-field Schr\"odinger equation~(\ref{eq.nlseq}) with additional Lindblad-type dissipative terms. It is, however, for the above reasons unclear whether this result is applicable to describe the evolution of mixed states of bosonic many-body systems. 

The difficulty to rigorously describe decoherence of single-particle observables of identical particles is entirely inherited from the particle's statistics. As is well known in the literature \cite{RevModPhys.35.668,Garrod,Kummer1967,Kummer1970,DavidsonBook,Radzki2010,RDMBook}, the symmetrization rules forbid the determination of a bosonic many-body state on the basis of its own single-particle distribution---unless that single-particle state is pure. In this case we obtain the pure product state~(\ref{eq.pureproductstate}) employed in the present paper. No solution at all can exist to this problem in the case of fermionic systems.

In conclusion, the \textit{coherent} pure-state dynamics of Bose-Einstein condensates can be microscopically described by the nonlinear mean-field Schr\"odinger equation~(\ref{eq.nlseq}). Its general form allows for the application to arbitrary systems of a single bosonic species with pairwise interactions. First steps towards the extension of our methods for the description of effectively \textit{incoherent} dynamics of systems of identical particles -- indubitably an equally important and difficult problem -- were discussed. Our results are limited mostly by the required assumption of a pure single-particle state at all times. Future work might aim at a combination of the present approach with a description of decoherence based on a stochastic pure-state evolution \cite{PhysRevLett.68.580,GardinerZoller,PhysRevA.63.023606,Carusotto2001,PhysRevA.69.022115}. Another possible extension of the present work could be achieved by dropping the projection onto a subspace of fixed particle numbers, e.g., by considering that the system is described, at all times, by a density matrix of an unspecified number of particles, $\rho\simeq\sum_{i=1}^N\rho^{(i)}$. This would also open up the possibility to describe incoherent particle loss or gain, which cannot be included in the present approach. Finally, the present approach could be combined with a tunable level of distinguishability by extending the description to additional internal degrees of freedom \cite{Ra2013,PhysRevA.91.013844,PhysRevA.91.022316}. This could be used to study the limit of the derived dynamical equations for distinguishable particles and their relationship with effective open-system descriptions, known from standard theory \cite{BreuerPetruccione2006}.

Possible applications of this theory could be in the context of transport dynamics with cold atomic systems, where interactions pose a significant limitation to the present theoretical understanding \cite{EPJ, PRA90, PRA93, Belzig, PRL117, Science1, Science2, PNAS}. In the fermionic experiments reported in Refs.~\cite{Science1, Science2, PNAS}, the formation of Cooper pairs or biatomic molecules leads to bosonic quasi-particles, which can be treated within the framework presented in this work.

\section*{Acknowledgments}
We thank Ugo Marzolino for frequent discussions during the early stages of this work. M.G. thanks the German Academic Scholarship Foundation (Studienstiftung des Deutschen Volkes) for support.


\end{document}